\documentclass[english,superscriptaddress]{revtex4}
\usepackage[T1]{fontenc}
\usepackage{color}
\usepackage{amsmath}
\usepackage{amssymb}
\usepackage{graphicx}
\usepackage{babel}

\usepackage{color}

\usepackage{times}

\begin{document}

\title{Optical Spectra of p-Doped PEDOT Nano-Aggregates Provide Insight into the Material Disorder}

\author{David Gelbwaser-Klimovsky}
\affiliation{Department of Chemistry and Chemical Biology, Harvard University, Cambridge, MA 02138}
\email{dgelbwaser@fas.harvard.edu}
\author{Semion K. Saikin}
\affiliation{Department of Chemistry and Chemical Biology, Harvard University, Cambridge, MA 02138}
\author{Randall H. Goldsmith}
\affiliation{Department of Chemistry, University of Wisconsin-Madison, Madison, WI 53706 }
\author{Al\'an Aspuru-Guzik}
\affiliation{Department of Chemistry and Chemical Biology, Harvard University, Cambridge, MA 02138}
\email{aspuru@chemistry.harvard.edu}

\begin{abstract}

Highly doped Poly(3,4-ethylenedioxythiophene) or PEDOT is a conductive polymer with a wide range of applications in energy conversion due to its ease of processing,  optical properties and high conductivity. The latter is influenced by  processing conditions, including formulation, annealing, and solvent treatment of the polymer, which also affects the polymer arrangement. Here we show that the analysis of the optical  spectra of PEDOT domains reveals  the nature and magnitude of the structural disorder in the material. In particular, the optical spectra of objects on individual domains can be used for the elucidation of the molecular disorder in oligomer arrangement which is a key factor affecting the conductivity.

\end{abstract}
\maketitle

%

their conventional inorganic counterparts in a number of  energy related applications \cite{sun2015review}, including  photovoltaics \cite{Facchetti_ChemMat2011,Lipomi_AdvMat2011}, fuel cells \cite{guo2012self,winther2008high} and low-cost and environmentally-safe thermoelectrics \cite{Dubey_JPolSci2011,Wang_AdvEnMat2015,Wei_Materials2015,Bubnova_NatMat2011}.

A central feature of many of these devices is a transparent, flexible conductor layer.  Among several materials employed for this layer, materials based on poly(3,4-ethylenedioxythiophene) (PEDOT) stand out due to their high hole conductivity, chemical stability, and transparency to visible light \cite{Groenendaal_AdvMat2000}. PEDOT-based structures have been incorporated in touch screens, light-emitting diodes, and photovoltaic elements, to mention a few examples \cite{Lovenich_PolSci2014}. Though multiple experimental investigations have addressed the molecular organization of PEDOT-based materials, the microscopic electronic states that lead to high conductance and the interplay between these states, optical properties, and the material structure has yet to be definitively characterized.  The main obstacle to elucidating this relationship is the strong structural disorder that appears on multiple length scales and is highly sensitive to the preparation procedure \cite{Groenendaal_AdvMat2000}.

In this work, we link the supramolecular packing of PEDOT with its optical absorption spectra. We argue that optical spectroscopy of PEDOT aggregates on a single domain level can be used for characterization of the microscopic electronic structure in the material that is a function of molecular disorder.  Specifically, we show how disorder in two packing models, widely accepted for PEDOT aggregates, leads to a transition between J- and H-aggregation \cite{spano2014h,siddiqui1999h} reflected in optical spectra. Finally, we identify the degree of disorder potentially responsible for this transition and study the changes in localization of the bright and dark states, in order to better understand the character of the spectral peaks. As we show below, disorder tends to localized the states and bright states are more delocalized than the dark ones. As a model system, we focus on  structures composed of multiply charged oligomers. This model matches the experimentally determined charge density in highly conducting PEDOT.

\begin{figure}[htbp]
\begin{center}
\includegraphics[scale=1.2]{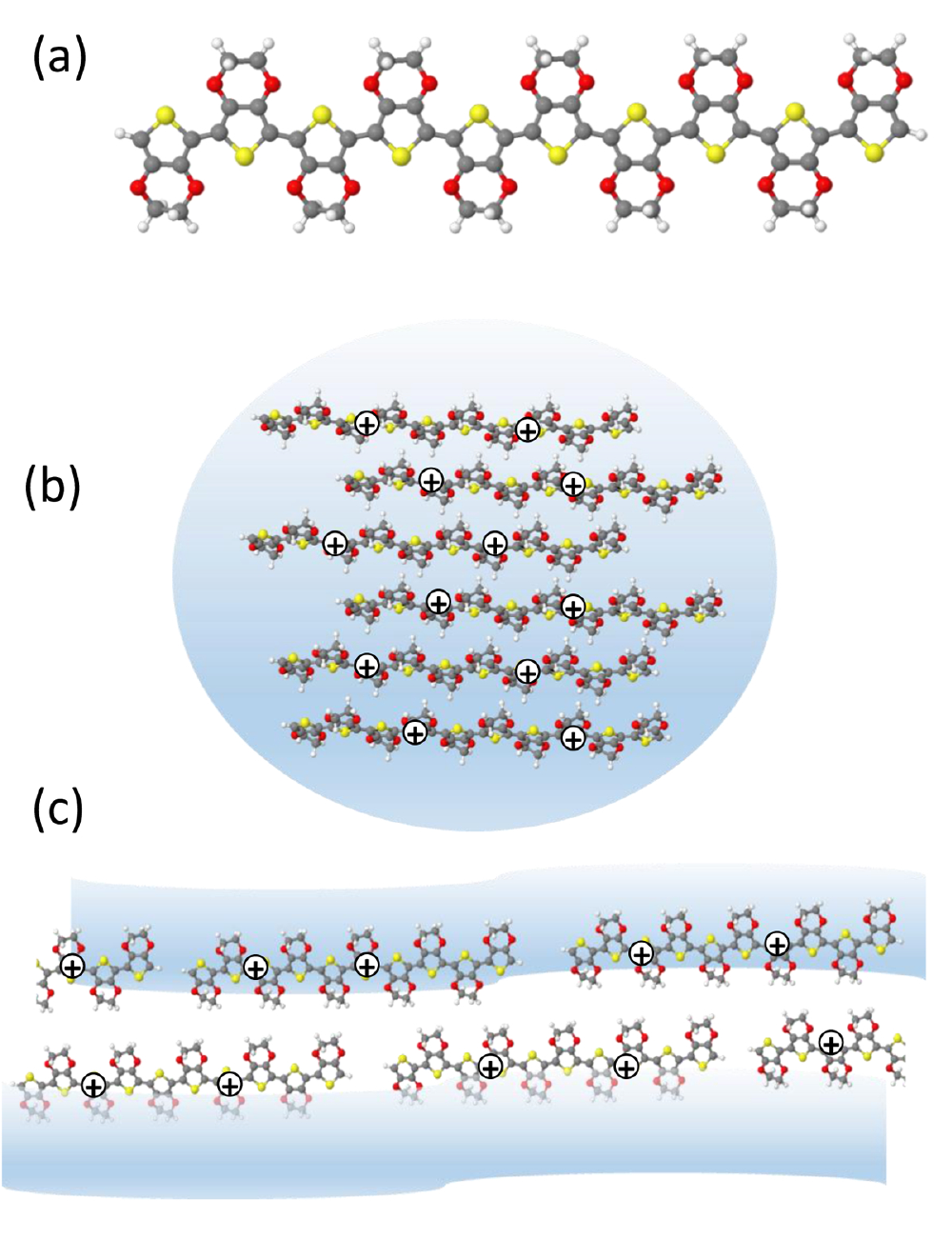}
\caption{PEDOT oligomer geometries (a) Optimized structure of a single PEDOT oligomer (some hydrogens not visible)  (b-c). Two packing models considered in the study: (b) a stack of parallel oligomers \cite{Dkhissi2008} and (c) a double-layer head-to-tail packing \cite{Lenz201144}. The blue ellipsoid and ribbon represent counterions equilibrating the PEDOT charges.}
\label{fig:struct}
\end{center}
\end{figure}

 A high-conductance regime is observed for p-doped PEDOT \cite{Groenendaal_AdvMat2003} when the injection of positive charges induces a structural relaxation effectively forming polaron and bipolaron states \cite{Bredas_AChR1985}. In this case, PEDOT-based materials consist of short oligomers, see Figure~\ref{fig:struct}(a), bound to counterions that balance the charges of oxidized PEDOT oligomers and affect the structural arrangement of the PEDOT oligomers. The most common forms of high-conducting PEDOT are PEDOT:tosylate and PEDOT:polystyrene sulfonate (PEDOT:PSS) \cite{PEDOT_book}. In the latter form the PSS polymer also plays the role of a structural backbone.
As compared to neutral oligomers, the ground state of p-doped oligomers has a quinoid form \cite{Bredas_CPL2002,Bredas_JACS2008}, which results in a delocalization of the charge carriers over several monomers. During deposition, typically via spincoating, the PEDOT oligomers rearrange, resulting in formation of supramolecular aggregates with a characteristic $\pi-\pi$-stacking intermolecular distance of $0.34$~nm.\cite{Takano_MacroM2012}

Figure~\ref{fig:struct}(b-c) shows two widely discussed packing motifs -- parallel stacks  \cite{Dkhissi2008} and head-to-tail chains \cite{Lenz201144}. Both motifs would exhibit similar $\pi-\pi$-stacking distances in x-ray scattering measurements. \cite{Takano_MacroM2012}. These general motifs have also been discussed in several variations including tilted stacking \cite{Bredas_JACS2008} and formation of 3D nanocrystals \cite{Takano_MacroM2012}. The physical reasoning behind the proposed packing begins with $\pi-\pi$ interactions that induce a parallel alignment of PEDOT oligomers and result in the formation of oligomer stacks. The electrostatic attraction to a backbone polyelectrolyte chain balances the self-repulsions from positive charge on the oligomers. This interaction tends to minimize the volume-to-surface ratio of the stacks, therefore limiting the stack size. The details of the deposition procedure, including solvent choice and post-processing, affect the balance between these two interactions and thus, controls the morphology of the structure \cite{Groenendaal_AdvMat2000, gasiorowski2013surface}.

In contrast with inorganic materials, the interactions mentioned above are relatively weak, frequently leading to high disorder in most conductive organic materials. This disorder influences both the optical spectra \cite{thiessen2013unraveling} and the charge transport properties  \cite{Friedlein_AdvElM2015}. For PEDOT-based materials, we can identify several specific types of disorder. First, the length and doping of each oligomer are not well controlled properties, resulting in a distribution of mesoscopic realizations and as a consequence, electronic states. Second, in a pair or stack of PEDOT oligomers, each individual oligomer can be displaced relative to its neighbor along the length of the oligomer. Finally, domains with  different packing structures, and with a different relative oligomer orientation, can be formed on a larger scale. These structural variations affect both the material conductance and optical spectra \cite{crispin2003conductivity,gasiorowski2013surface}.

Heavily doped PEDOT is mostly transparent to visible light. However, it has a very broad and structureless absorption spectrum below 1~eV \cite{Massonnet2014}. This spectral feature is associated with the HOMO-1 to HOMO and also HOMO to LUMO transitions of cation oligomers \cite{Bredas_AChR1985}.
As compared to isolated PEDOT oligomers, the absorption spectra of PEDOT aggregates are modified by the Coulomb coupling between molecular excitations (excitonic interactions) \cite{forster1948zwischenmolekulare,Foe65_93,Dexter_JCP1953} as well as by charge tunneling \cite{Osterbacka839}. Depending on the relative alignment of the molecules, the excitonic interactions can result in the shift of the absorption peak to longer (J-aggregation) or shorter (H-aggregation) wavelengths \cite{spano2014h,SKS_NanoPh2013}. Many examples of J-aggregates can be found in nature, where aggregates of pigments play an essential role in light-harvesting and energy transfer in phototrophic organisms \cite{Green_book,Croce_NatChemBiol2014, Huh_JACS}.

To predict and characterize optical properties of PEDOT aggregates we use a theoretical model, where single oligomer properties are computed using Time-Dependent Density Functional Theory (TDDFT), while the emergent aggregate spectra are calculated based on a Hubbard-type model \cite{valleau2012exciton}. Details of TDDFT calculations are provided in the Supplementary Information. As compared to previous theoretical studies of PEDOT optical spectra \cite{Lenz201144,Gangopadhyay_RCS_Adv2014, Dkhissi2008}, here we focus on the formation of small disordered clusters that include random displacements rather than a periodic crystalline orientation or arrangement of PEDOT oligomers on a polymer backbone. To our knowledge, the role of structural disorder in the optical properties of PEDOT-based materials has not yet been explored in detail.

The formation of PEDOT aggregates is driven by inter-oligomer forces. $\pi$-$\pi$ interactions between the aromatic oligomers orients them face-to-face and fixes the perpendicular distance to $0.34$~nm  as determined by X-ray scattering \cite{Takano_MacroM2012,aasmundtveit1999structure,atanasov2014highly}. The energetic benefit of $\pi$-orbital overlap makes it energetically favorable for neighboring monomers to be cofacially aligned in a way that maximizes orbital overlap. Using this tendency as a simplifying assumption, in our models we constrain the relative parallel displacement between oligomers  to multiples  of the monomer length. Below, we analyze the packing motifs shown in figs. \ref{fig:struct}(b-c). These models are then used as building blocks for more complicated structures, e.g. 2D films \cite{martin2010morphology} or 3D nanocrystals \cite{Takano_MacroM2012,cho2014single}.

The analyzed structures are composed of around 100 oligomers, which is enough to saturate the spectral shifts related to the molecular arrangement size. In the cases where disorder is allowed, we average over 10,000 realizations, which is sufficient to obtain a wide sampling of the possible noise realizations. Considering larger structures or  averaging over more realizations introduces only minor  changes in  the results.

\begin{figure*}[hb]
\begin{center}
\includegraphics[width=\textwidth]{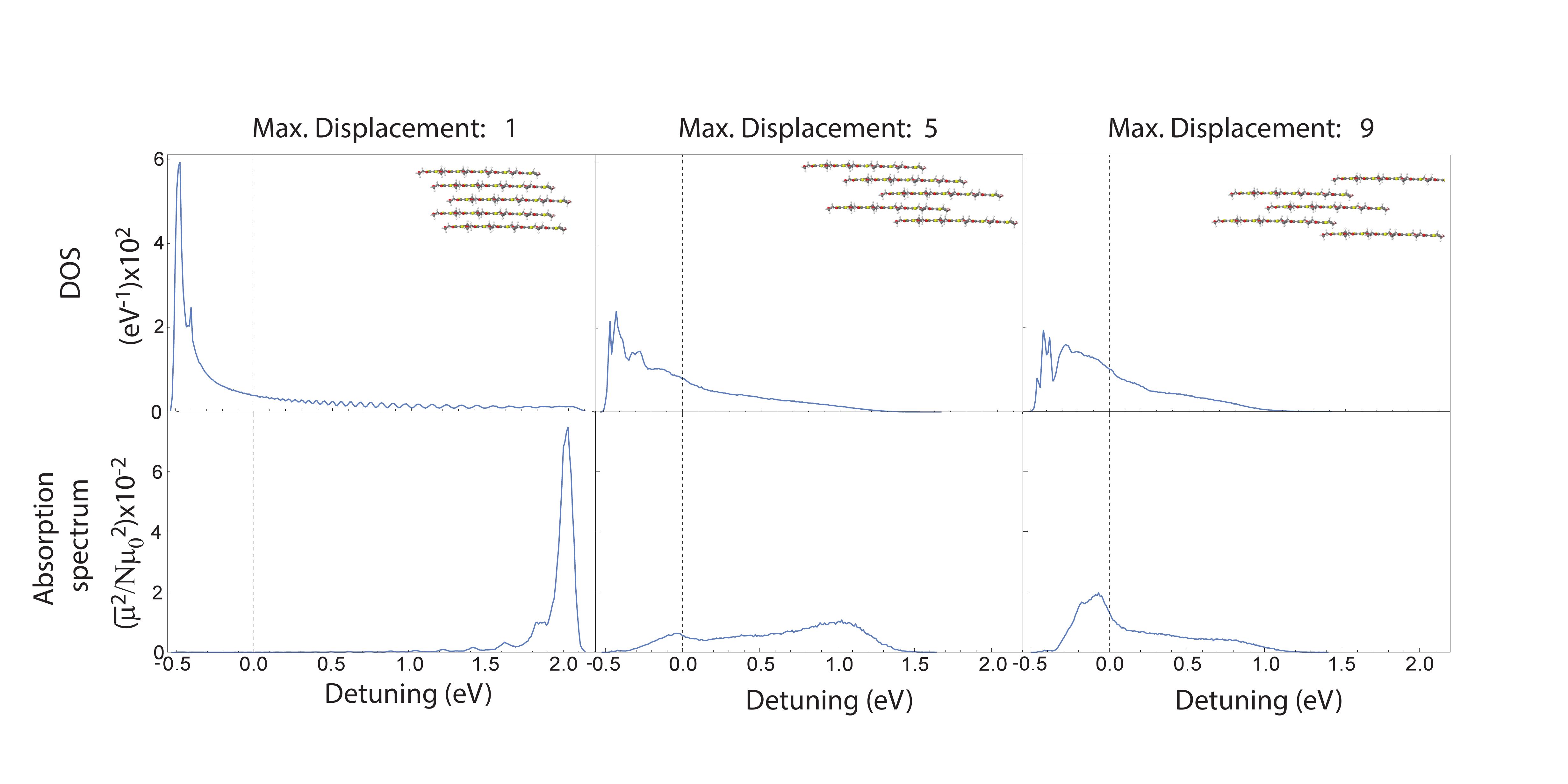}
\caption{Average of DOS (top) and absorption spectra (bottom) over 10,000  random realizations of single chain arrangements as function of the
detuning.  Zero detuning is the  single oligomer transition energy. From left to right we raise the disorder level by increasing the maximum allowed parallel displacement (1,5,9 from left to right). By enlarging the maximum  displacement, see insets, the number of possible realizations is escalated and with it the disorder level. This is reflected in the DOS width which grows with the level of disorder. On the other hand the spectra experience a red shifting due to the change in the structural arrangement (see the main text). Even in the case with the largest red shift, right plots, the lowest energy states of the DOS are not bright. Insets: Examples of  small parts of  realizations of the corresponding disordered single oligomer stack.}
\label{fig:singledos}
\end{center}
\end{figure*}

\begin{figure}
\centering{}\includegraphics[width=\textwidth]{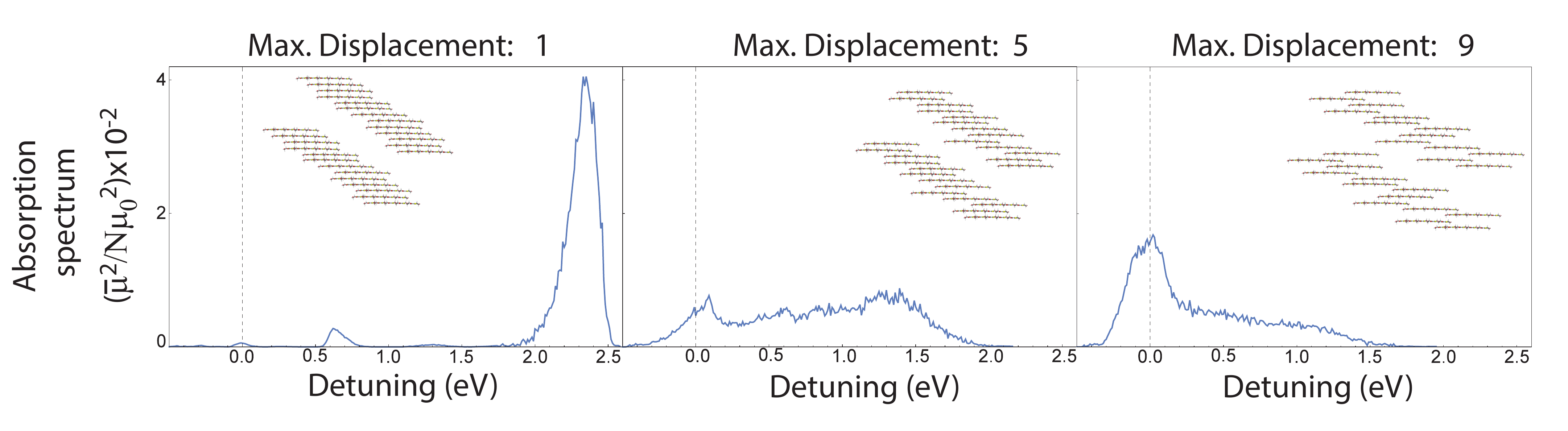}\caption{Average spectra of 10,000 random realizations of the structural arrangement found in Ref.  \citenum{Takano_MacroM2012} as function of the
detuning. This structure is composed of two parallel 13-oligomer chains. From left to right we rise the disorder level by increasing the maximum allowed parallel displacement (1,5,9 from left to right). As a simplifying assumption we assume that the two 13 oligomer chains have the same disorder realization, therefore they are identical. Insets:  Examples of corresponding realizations of the structural arrangement proposed in  Ref. \citenum{Takano_MacroM2012}, with disorder included  in the form of random parallel displacement between oligomers.}
\label{fig:talspe}
\end{figure}

\textit{Parallel Stack of Oligomers.}
As a first case, we studied a cofacially stacked column of 100 parallel oligomers  as shown on \ref{fig:struct}b. Perfectly aligned oligomers form an H-aggregate, with a blue-shifted spectrum relative to the single isolated oligomer. The introduction of a constant parallel displacement between oligomers results in a transition from an H-aggregate behavior to J-aggregate behavior,\cite{valleau2012exciton} see also Supplementary Information for more details. This change is reflected in the initial blue-shift reduction, which eventually turns into a red shift relative to the isolated oligomer as the constant displacement between neighboring oligomers is further enlarged. Nevertheless, even at the maximum possible parallel displacement that maintains at least one monomer overlap, the lowest states do not contribute to the absorption spectra, in contrast to the case of a perfect J-aggregate. We also analyzed the effect of  disordered or random parallel displacements that  keep the monomers aligned and with at least a one monomer overlap. Each realization consists of a disordered chain composed of 100 parallel oligomers, with a random parallel displacement between them.  Figure ~\ref{fig:singledos} shows the density of states (DOS) and absorption spectra averaged over 10,000 realizations. To mimic structural disorder, we allow random displacements along the long axis of the oligomer over a range of amplitudes in both directions (forward or backward) up to a maximum value. The maximum value for the lowest disorder level is a single monomer displacement. The low disorder level is reflected in the small widths of the DOS and the absorption spectra (see Figure \ref{fig:singledos}) as wells as the spectral blue shift relative to the isolated oligomer. A low disorder chain is similar to the ordered case, with parallel arrangement resulting in a blue-shifted spectra. The disorder level grows by increasing the maximum parallel displacement (see Figure \ref{fig:singledos}-Insets).  Larger disorder results in broader  spectra and DOS, Figure \ref{fig:singledos}. The increase in the amplitude of random parallel displacements transforms the arrangement into a mixture of a parallel and a head-to-tail chain. In turn, this reduces the spectral blue-shift relative to the isolated oligomer, eventually becoming a red-shifted spectra as the maximum allowed displacement is further increased. In spite of this, these arrangements also fail to form a perfect J-aggregate, because the lowest  states do not participate in the absorption spectra. The spectral shifting behavior also arises in more complex models. Delocalization analysis clarifies the nature of the intense bands of the absorption spectra.  In particular, we calculate the inverse participation ratio \cite{smyth2012measures} and find that the bright states are consistently more delocalized than the dark states, independent of the oligomer displacement (see Supplementary Information). This behavior also persists in more structurally complicated models. Nevertheless, increasing the degree of disorder increases the localization independently of whether the state is dark or bright, as expected. To test  our approach, we used the structural arrangement proposed in  Ref. \citenum{Takano_MacroM2012} for the PEDOT:PSS micelles in a solid film, where the PEDOT oligomers form nanocrystals in the micelle core  surrounded by PSS counterions. The PEDOT oligomers are arranged in two side-by-side stacks, where each stack is composed of a 13 cofacial oligomers. Figure ~\ref{fig:talspe} shows the average spectra for the PEDOT nanocrystal for different disorder levels. Similar to the previous model,  augmenting the maximum parallel displacement, the spectrum is red shifted. The peak in the absorption spectrum of the most ordered case (maximum displacement of a single monomer) at around 0.6 eV is a consequence of the short length of the chains. For a completely ordered aggregate, this peak corresponds to the next odd delocalized excitation after the brightest state.

\textit{Head to tail chains}
We next considered an arrangement consisting of two parallel chains \cite{Lenz201144} as shown in Figure \ref{fig:struct}c. Each chain is composed of 50 oligomers in a head-to-tail arrangement.  The two chains are cofacially stacked upon each other. The minimal inter-oligomer distance within a chain is limited by the Van der Waals radius, and the interchain distance is $0.34$~nm, the same as in the previous model.

\begin{figure}
\centering{}\label{doublechain}\includegraphics[width=1\textwidth]{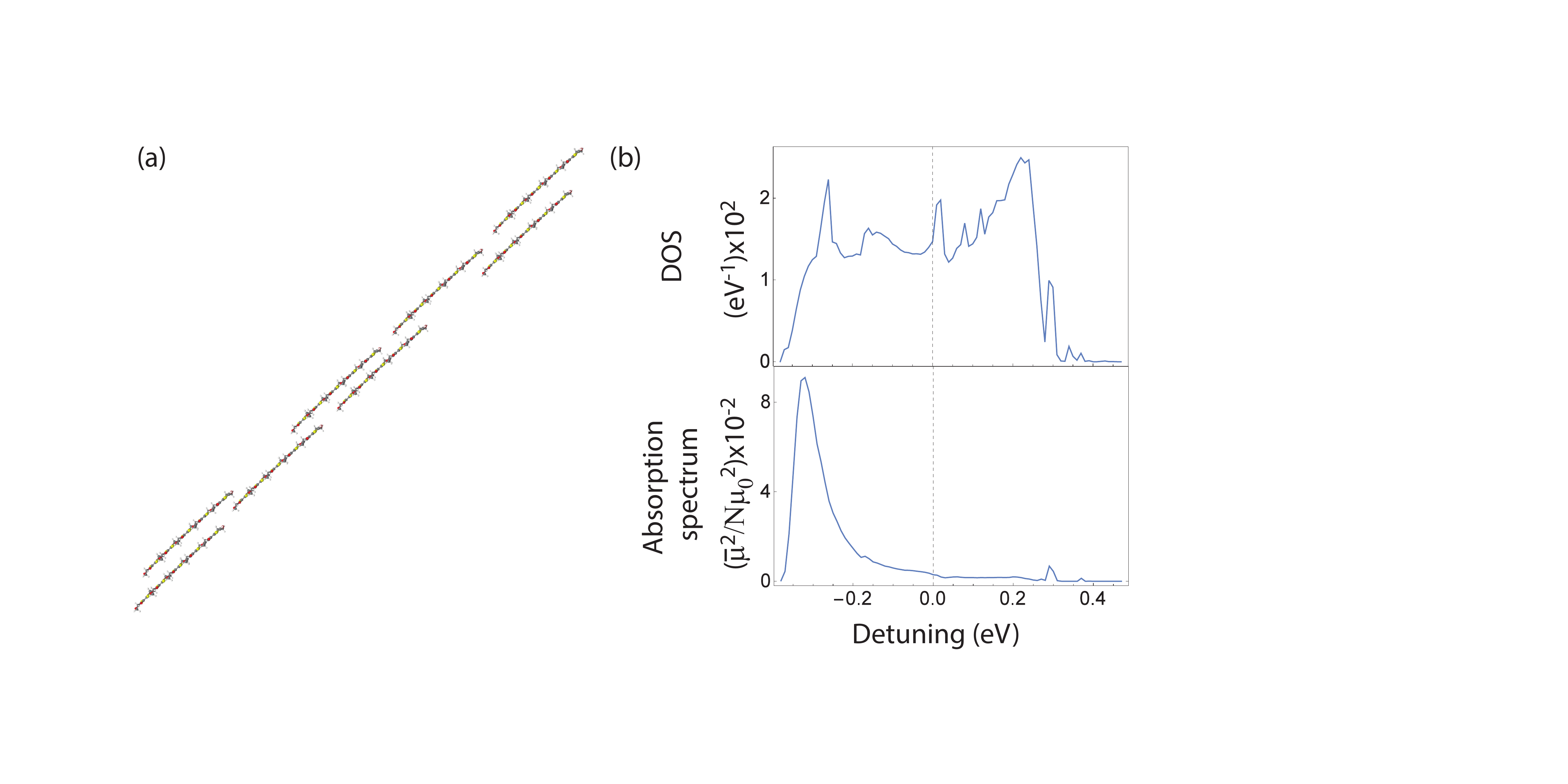}\caption{a) Part of a disordered double chain, shown for illustration purpose (the complete double chain is formed of 100 oligomers). b) Average DOS (top) and spectra (bottom) over 10,000  realizations.   In contrast to the other structural arrangements, the lowest energy states of the DOS are bright.}
\label{fig:doublechain}
\end{figure}

We first considered two identical and perfectly aligned chains with a minimum distance between oligomers. The absorption spectrum is blue shifted relative to the single monomer spectrum. As we displaced one chain relative to the other, the blue-shifting is reduced, eventually evolving into a red-shifted spectrum, as seen above. A more complex model allows  structural disorder due to a random degree of overlap between oligomers in the two cofacially stacked chains that preserves the following constraints: i) Facing monomers are aligned; ii) the minimal head to tail distance is $2.5\dot{A}$; iii) chain continuity is ensured by requiring that every oligomer  overlaps with two oligomers from the second chain. The randomness of alignment allows several arrangements
 and the results shown are the average over 10,000 realizations (see Figure \ref{fig:doublechain}). In contrast to the single cofacial stack of oligomers, the red shifting for the head to tail pair of chains is larger and more narrow. Also here, the lowest  states contribute to the absorption spectra, in resemblance of a perfect J-aggregate. As in the single cofacial stack of oligomers, the bright states are delocalized while the dark are more localized  (see Supplementary Information.)

%
%
\begin{figure}[h]
\centering{}\includegraphics[width=1\textwidth]{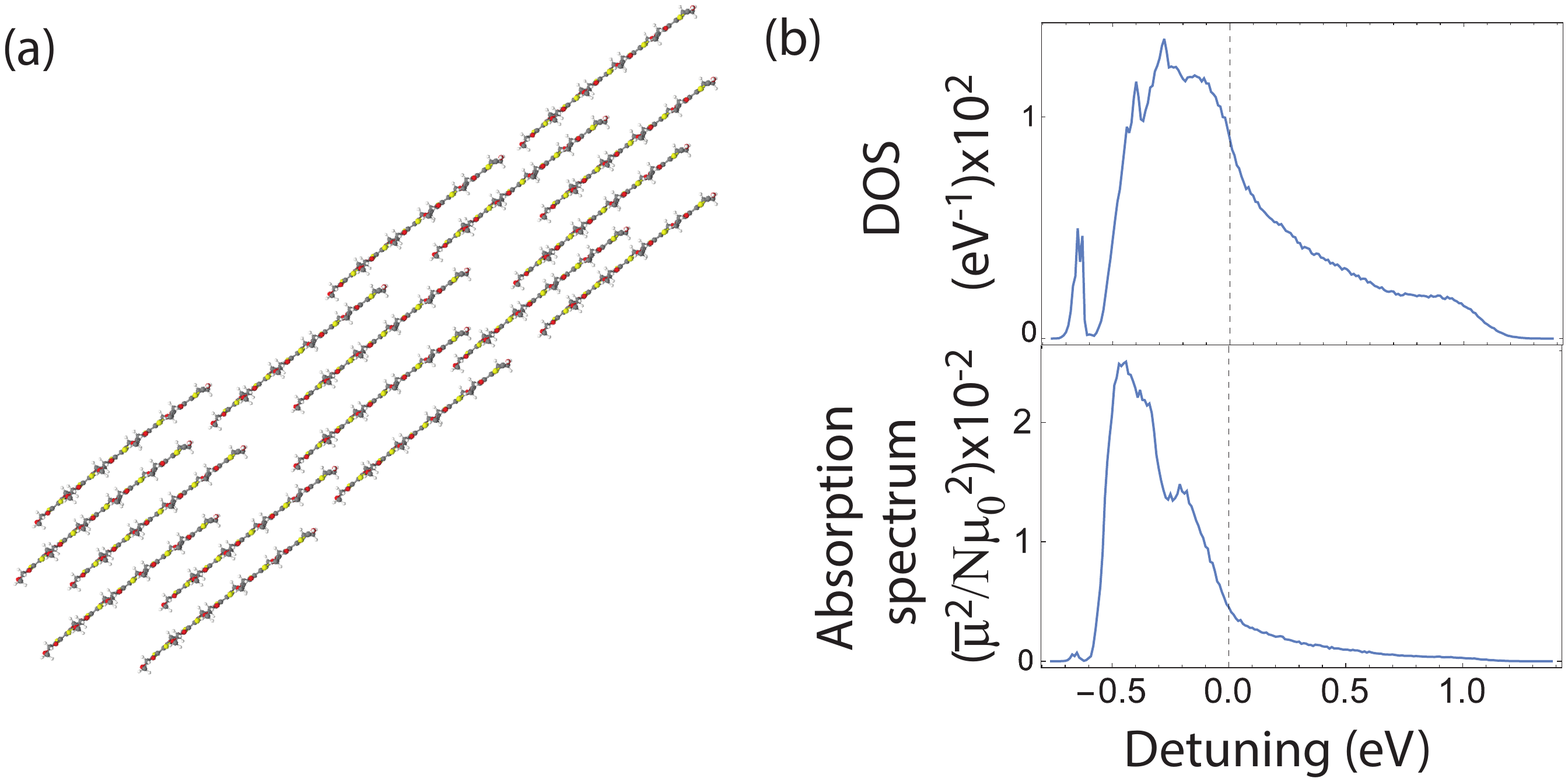}\caption{2D layer structure (a) and average DOS (top-b) and spectrum (bottom-b) over 10,000 realizations of the disordered 2D layer. There is a  low energy peak in the DOS, due to the $\pi$-$\pi$ stacking of head-to-tail chains. Here again the lowest energy states of the DOS are dark.}
\label{fig:2dlayer}
\end{figure}

The insights obtained from the two basic models introduced earlier can help in the examination of the electronic structure of more elaborate arrangements. In order to demonstrate these insights we analyzed a 2D layer, as the one proposed in Ref. \citenum{martin2010morphology} for the condensed state of a vapor-phase polymerized PEDOT, where a 2D layer of PEDOT is ``sandwiched'' by  2D layers of PSS counterions. The structure of the PEDOT layer is composed of pairs of head-to-tail chains. The arrangement of the effective dipoles in the 2D layer is shown in Figure \ref{fig:2dlayer}. Due to the disorder, nearest neighbors may be relatively far away from each other, therefore only extended dipole-dipole interactions are considered. Each 2D layer is composed of 95 oligomers. The DOS and absorption spectrum are shown in figure \ref{fig:2dlayer}. This structure can be understood as a combination of the parallel oligomers chain and the head-to-tail chains models described above. In contrast to previous models, there is a low energy  peak in the DOS. The origin of this red shifted peak is the $\pi-\pi$ stacking of more
 head-to-tail chains that augments the total interaction, red shifting part of the DOS. In agreement with our results on parallel stacks of oligomers above, the lowest states do not contribute to the absorption spectra. This example shows how the intuitions obtained from simple models can be combined to analyze more intricate spectra.

A common feature of the studied models is the significant changes on the DOS and absorption spectra caused by the introduction of structural disorder. Furthermore, we showed that structural disorder is needed for reproducing the red shifted or low energy absorption spectra seen experimentally. Nevertheless, in general the lowest states  do not participate in the absorption spectra. The only exception is the disordered head to tail stack of oligomers, which becomes a perfect J-aggregate in the sense that the lowest states contribute to the energy absorption (see Figure \ref{fig:doublechain}).

   Thin films of PEDOT:PSS show a broad, featureless absorption spectrum below 1 eV \cite{Massonnet2014}.  This spectral behavior is in contrast to that of the radical cations and dications in phenyl-capped PEDOT oligomers, as demonstrated via solution-phase spectroelectrochemistry \cite{martin2000monodisperse}.  These solution phase measurements can be viewed as representative of the absorption spectra of isolated, un-aggregated PEDOT oligomers, essentially a starting point for building a thin film from a molecular building block.  Our calculations provide a guide to predict the thin film DOS and optical properties from the properties of the component molecules.  Absorption features of the phenyl-capped oligomers showed the expected $1/N$ length dependence, and extrapolation of dication absorption features of shorter oligomers to the oligomers of length ten considered above suggest a bipolaron absorption feature at ~1.2 eV.  If disorder is not included, a net blue shift of 1-2 eV is predicted due to the Coulomb interaction, in stark contrast to the experimentally observed thin film spectra.  In contrast, inclusion of disorder faithfully reproduces a red shift of bipolaron spectral features, suggesting broad absorption at 1 eV and below in the thin film, as observed.

	In conclusion,	we have shown the magnitude of structural disorder in the overlap between neighboring PEDOT oligomers in pi-pi stacked geometries has a significant impact on electronic absorption spectra.  Specifically, the experimentally observed red-shifted absorption spectra relative to isolated PEDOT oligomers could only be obtained if structural order was introduced above a critical displacement. Therefore we conclude that in PEDOT:PSS films, structural disorder is significant and has important contributions to both optical and electronic properties.  A variety of structural models were considered and all showed the need for structural disorder in order to reproduce red-shifted spectra.
		Finally, we point out that though X-ray scattering has provided important clues as to the nature of molecular packing in PEDOT materials, scattering measurements alone cannot be used to differentiate between different models of structural disorder.  Electronic spectroscopy on whole films can provide additional clues, but the significant heterogeneous broadening of spectral features limits the utility of optical spectra.  However, single-particle spectroscopy \cite{heylman2014photothermal,knapper2016chip} on individual polymer domains \cite{barbara2005single,thiessen2013unraveling}  can potentially add significant new information on the structural and electronic properties of the PEDOT family of materials.

\section{Acknowledgments}
We acknowledge the support from the Center for Excitonics, an Energy Frontier Research Center funded by the U.S. Department of Energy under award DE-SC0001088  (D. G.-K.,S.S and A. A.-G.)  and from the National Science Foundation under award DMR-1610345 (R.H.G.). We appreciate useful discussion with Dr. Dmitrij Rappoport and Dr. Dmitry Zubarev. First principles computations were run on  Odyssey cluster of the Harvard University, supported by the Research Computing Group of the FAS Division of Science.


\newpage

\section{Optical Spectra of p-Doped PEDOT Nano-Aggregates Provide Insight into the Material Disorder:
Supplementary Information}

\begin{figure}
\begin{center}
\includegraphics[scale=1.2]{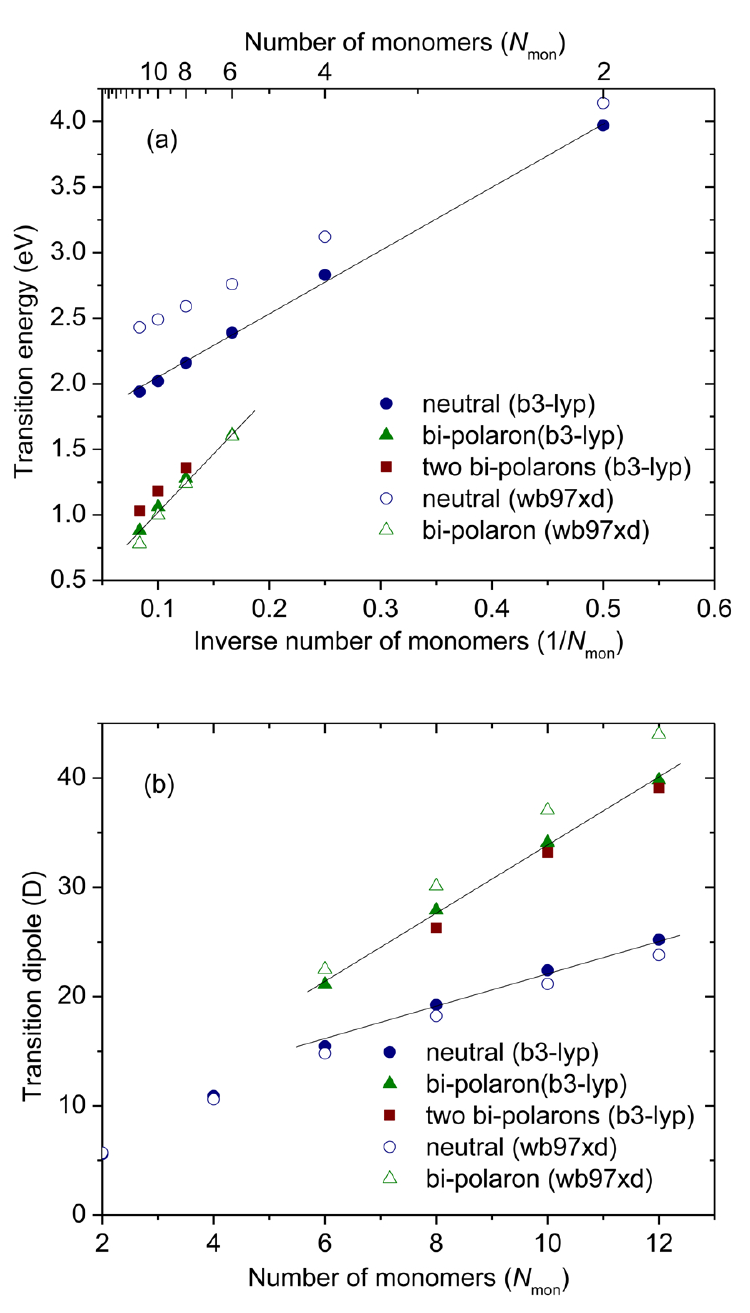}
\caption{Computed transition energies (a) and the transition dipoles (b) of neutral and double-charged PEDOT oligomers (bi-polaron and two bi-polaron) as functions of the oligomer length.}
\label{fig:mon_scale}
\end{center}
\end{figure}

For the TDDFT calculations of single oligomer properties, we used the quantum chemistry package Turbomole, version 6.0. \cite{TM_general}. The geometry optimization was done using a triple-$\zeta$ valence-polarization basis set (def2-TZVP \cite{BS_def2}) and B3LYP hybrid functional \cite{B3LYP}. In order to account for the neutral, polaron and bi-polaron ground states of the oligomers, the structures were optimized with the charges $Q=0,+1,+2,$ respectively. We also computed several quadruple-charged $n=4$ oligomer structures. The effects of the polarizable medium were accounted for using the conductor-like solvation model (COSMO) with effective dielectric constant $\epsilon=4$.\footnote{Variation of $\epsilon$ between 1 and 4 results in a small shift in transition frequencies.} The optimized structures are provided as edt$N$\_$n$.xyz files, where $N$ is the number of monomers and $n$ is the charge. The five lowest electronic excitations were computed for each oligomer. The influence of the oligomer length on the spectra of electronic excitations was studied in a set of oligomers of size $N_{\rm mon}=2,4,6,8,10$ and $12$ units.  We observe a strong electronic transition for neutral and double charged oligomers longer than 4 units, and for quadruple charged oligomers longer than 6 monomer units. This transition occurs mostly between the HOMO and LUMO orbitals of the oligomers. The examined structures span the experimentally determined estimates of oligomer length (6-18 monomer units) \cite{Kirchmeyer_JMChem2005} and charge density (4 monomers/polaron) \cite{crispin2003conductivity}. For the studied oligomer lengths, the energy of the transition scales as $1/N_{\rm mon}$, see Figure~\ref{fig:mon_scale}(a), similar to the expected dependence of a particle-in-a-box model assuming that the electron state is delocalized over the conjugated chain \cite{ANGE:ANGE19590710302, Bar_JCP1960}. Similar $1/N_{\rm mon}$ dependencies have been observed in a variety of experimental investigations of optical properties of solution-phase (isolated) polymers \cite{Bredas_JACS1983,Tolbert_AcChemRes1992,Apperloo_ChemEurJ2002}. In Table~\ref{tbl:prop} we collected the total energies, energies of the lowest electronic excitations, oscillator strengths associated with these excitations, and the leading contribution of the molecular orbitals for neutral and doubly charged oligomers.

\begin{table}
\begin{center}
\caption{Properties of neutral, and double-charged PEDOT oligomers computed with TDDFT (B3LYP functional): number of monomers,$N_{\rm mon}$, oligomer charge, $Q$, ground state energy,$E_{\rm g}$, energy of the lowest strong electronic excitation, $E_{\rm ex}$, and the contribution of HOMO-to-LUMO transition into these excitations.}\label{tbl:prop}
\begin{tabular}{|l|c|c|c|c|c|}
\hline
$N_{\rm mon}$ & $Q$ [e] & $E_{\rm g}$ [Hartree] &  $E_{\rm ex}$ [eV] & Osc. strength &  Orbitals \\
\hline
2 & 0 & -1559.716 & 3.97 & 0.47 & 98\% \\
4 & 0 & -3118.261 & 2.83 & 1.28 & 99\% \\
6 & 0 & -4676.784 & 2.39 & 2.17 & 98\% \\
6 & 2 & -4676.415 & 1.61 & 2.72 & 99\% \\
8 & 0 & -6235.318 & 2.16 & 3.03 & 97\% \\
8 & 2 & -6234.965 & 1.28 & 3.81 & 99\% \\
10 & 0 & -7793.881 & 2.01 & 3.85 & 95\% \\
10 & 2 & -7793.565 & 1.06 & 4.66 & 98\% \\
12 & 0 & -9352.386 & 1.94 & 4.67 & 93\% \\
12 & 2 & -9352.049 & 0.88 & 5.29 & 98\% \\
\hline
\end{tabular}
\end{center}
\end{table}

\begin{figure}
\begin{center}
\includegraphics[scale=1.5]{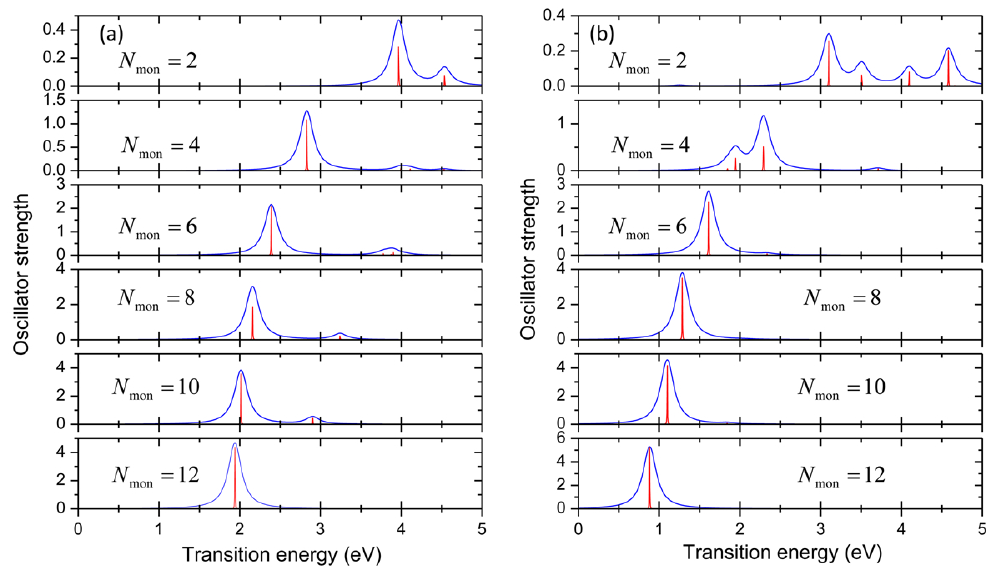}
\caption{Electronic excitation spectra of (a) neutral and (b) double-charged PEDOT oligomers (bi-polaron) composed of $N_{\rm mon}$ monomers. The transition frequencies within the $0-5$~eV range, red lines, are computed using TDDFT, B3LYP functional. The empirically broadened spectra with a $100$~meV linewidth are show in blue.}
\label{fig:mon_spectra}
\end{center}
\end{figure}

As the oligomers are allowed to increase in size the absorption peak saturates at about $1.5$~eV for the neutral oligomers, and the double and quadruple charged molecules exhibit transition frequencies significantly below $1$~eV. In Figure~\ref{fig:mon_spectra} we show computed spectra of six oligomers with net charges $Q=0,+2$. As demonstrated in previous calculations, \cite{Bredas_AChR1985} the single-charged chains exhibit two transitions, a weak transition composed mostly of HOMO-1 - HOMO orbitals in the sub-eV range and a stronger HOMO - LUMO transition (data not shown). To verify that a B3LYP hybrid functional gives correct estimates for delocalized electronic excitations in the studied oligomers, \cite{Nayyar_JPCL2011,Salzner_JCTC2011} we also computed the excitation energies and the transition dipoles with a range-separated functional wB97XD  \cite{Chai_PCCP2008} and the basis set 6-31+G* as implemented in Gaussian~09 \cite{g09}. The computed results (open circles and open triangles in Figure~\ref{fig:mon_scale}), show linear trends similar to the ones obtained for B3LYP with a systematic shift.  In particular, the transition energies of the bipolarons, the critical species in this work, show little difference between the two functionals. We expect that the deviations due to long-range corrections\cite{Nayyar_JPCL2011} will be more pronounced for longer oligomers, but this work only considered the short oligomers that are present in PEDOT:PSS.

\begin{figure}
\begin{center}
\includegraphics[scale=0.7]{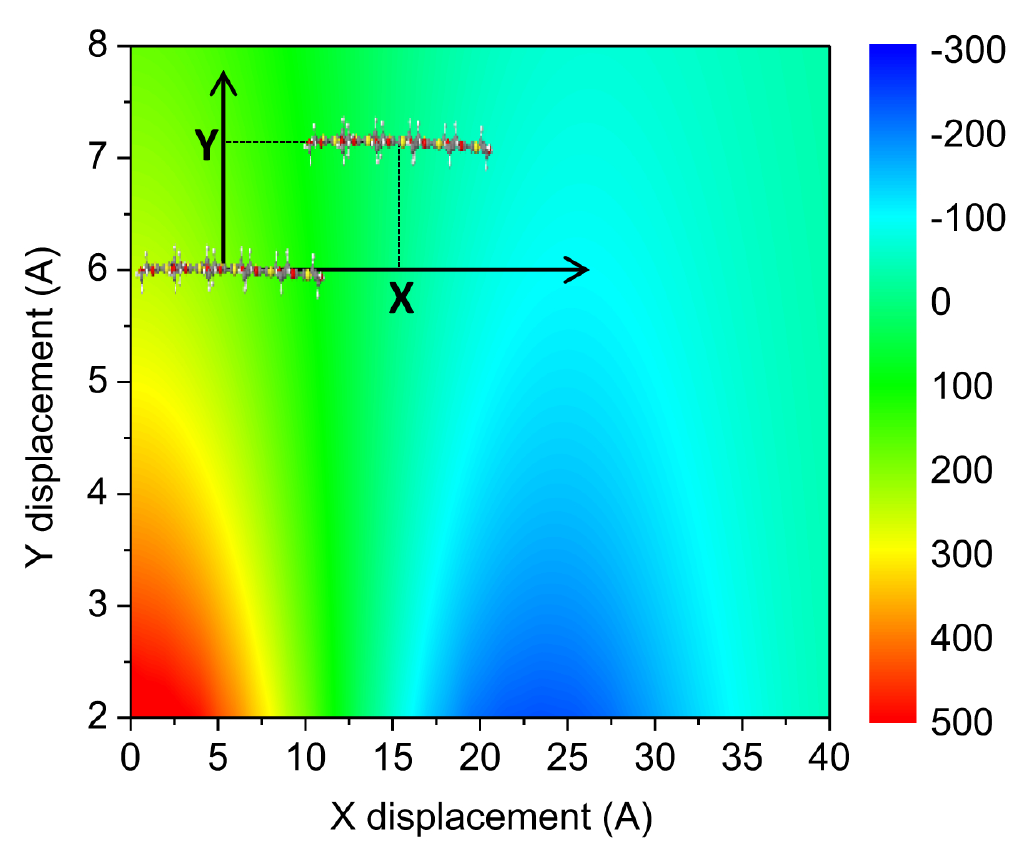}
\caption{A 2D map of the F\"orster interaction between two 10-monomer PEDOT chains parallel displaced relative to each other (see inset). The axis values refer to center-center distances. The color scheme shows the intermolecular coupling in meV.}
\label{fig:F_map}
\end{center}
\end{figure}

The electronic excitation spectra of molecular aggregates are significantly modified as compared to the isolated molecules that compose these aggregates. The main sources of the modifications are the Coulomb coupling between electronic excitations, including both the F\"orster \cite{Foe65_93} and the Dexter exchange \cite{Dexter_JCP1953} interactions, formation of charge-transfer excitations or delocalized hole bands \cite{Osterbacka839}, and changes in the dielectric properties of the local environment. The Coulomb interaction between the transition dipole moments, specifically the F\"orster coupling, can be strong in PEDOT aggregates.  The transition dipole scales approximately linearly with the number of monomers, reflecting a delocalization length of the electronic excitations over the length of the oligomer, see Figure~\ref{fig:mon_scale}(b). For a double-charged molecule composed of ten monomers the value of the transition dipole is about $30$~Debye.  For a more generalized characterization of the Coulomb coupling between a pair of oligomers we computed the F\"orster and Dexter couplings as
\begin{align}
    V_{\rm F}&= \int \mathrm{d}^3x \int \mathrm{d}^3y \, \psi_{\rm H_1}(x) \psi_{\rm L_2}(y) \frac{e^2}{|x-y|}\psi_{\rm L_1}(x) \psi_{\rm H_2}(y), \label{V_F}\\
    V_{\rm D}&= \int \mathrm{d}^3x \int \mathrm{d}^3y \, \psi_{\rm L_2}(x) \psi_{\rm H_1}(y) \frac{e^2}{|x-y|}\psi_{\rm L_1}(x) \psi_{\rm H_2}(y),\label{V_D}
\end{align}
\noindent where the indices $\rm H_{i}$ and $\rm L_{i}$ correspond to the HOMO and LUMO orbitals of an $i$-th molecule \cite{valleau2012exciton}. Equation \ref{V_F} describes the F\"orster interaction beyond the dipole approximation as a Coulomb transition between the initial $\psi_{\rm L_1}(x) \psi_{\rm H_2}(y)$ and the final $\psi_{\rm H_1}(x) \psi_{\rm L_2}(y)$ states of a two-oligomer system. In Equation \ref{V_D}, the transition involves tunneling of electrons between oligomers, and this process requires an overlap of electronic wavefunctions. For all intermolecular distances relevant to our system, the Dexter interaction between the molecules is less than $5\%$ of the F\"orster coupling. Figure~\ref{fig:F_map} shows a map of the F\"orster coupling as a function of the relative spatial positions of two 10-monomer oligomers. The negative sign of the F\"orster coupling corresponds to the J-aggregation of the molecules, while the positive sign of the coupling indicates H-aggregation, with the sign of the coupling directly linked to the expected ordering and symmetry of the new J- or H-aggregated hybrid orbitals. In general, if the oligomers are aligned in parallel the excitonic coupling between the oligomers leads to a blue-shift of the collective absorption peak. Only oligomers with a large relative displacement (more than 2/3 of the oligomer length) along the x-axis show J-aggregation in the absorption spectra, with a consequent red-shift of the absorption peak. It should be noted that due to delocalization of the electronic excitations the nearest neighbor F\"orster interaction between oligomers decreases with the length of the chains,\cite{Manas_JCP1998, Barford_JCP2007} which may result in reduced F\"orster coupling for sufficiently long chains. We observe this effect qualitatively, with a 10-monomer chain showing an interaction subtly reduced from the 8-monomer chain.  However, at the short oligomer lengths determined in experimental characterizations of PEDOT:PSS, we expect this effect to have only a minor effect.

In order to accelerate the computation of the electronic excitation spectra of aggregated oligomers, we treated the F\"orster coupling beyond nearest neighbor interactions with an extended dipole model \cite{Ko96__,valleau2012exciton}
\begin{align}
V_{\rm F}^{\rm ed}=k_e q^{2}\left(\frac{1}{r_{++}}+\frac{1}{r_{--}}-\frac{1}{r_{+-}}-\frac{1}{r_{-+}}\right),
\label{V_F_ed}
\end{align}
\noindent where $q$ is the effective dipole charge and $r_{ij}$ is the distance between the $i$ charge on one oligomer and the $j$ charge on the another oligomer and $k_e$ is the Coulomb constant. The F\"orster interaction 2D map is used to calibrate the effective dipole charge as well as its length. This information together with a model of the structural arrangement of the oligomers is enough to construct the collective Hamiltonian and the electronic absorption spectrum \cite{valleau2012exciton}.

Charge transfer  (CT) excitations in conducting polymers have been observed previously \cite{Osterbacka839} and analyzed theoretically \cite{Beljonne_AdvFM2001}. Usually, the CT absorption line is red-shifted as compared to the intramolecular excitations. These transitions are allowed if there is a sufficient overlap between the electron wavefunctions of different oligomers. The lowest order term contributing to the CT transition in an oligomer dimer is due to the excitation from a HOMO orbital of one oligomer to a LUMO orbital of another oligomer. For two parallel-stacked 10-monomer units displaced by $0.34$~nm the CT transition dipole, calculated as a matrix element of a dipole operator taken between HOMO and LUMO orbitals of adjacent oligomers, the interaction strength is $\mu_{\rm CT}=1.4$~D. This value is about one order of magnitude smaller that the transition dipoles for intramolecular excitations. Previous theoretical studies of delocalized electronics states in P3HT also showed that the intensities of charge-transfer excitations are small compared to the intramolecular excitations \cite{Beljonne_AdvFM2001}. While we expect that the far infrared part of the absorption spectra of PEDOT aggregates will have significant contribution from CT excitations, the near infrared and visible parts of the absorption spectrum can be largely attributed to the Coulomb interaction between the oligomers.

\section*{States delocalization}
\begin{figure}
	\centering
		\includegraphics[width=1\textwidth]{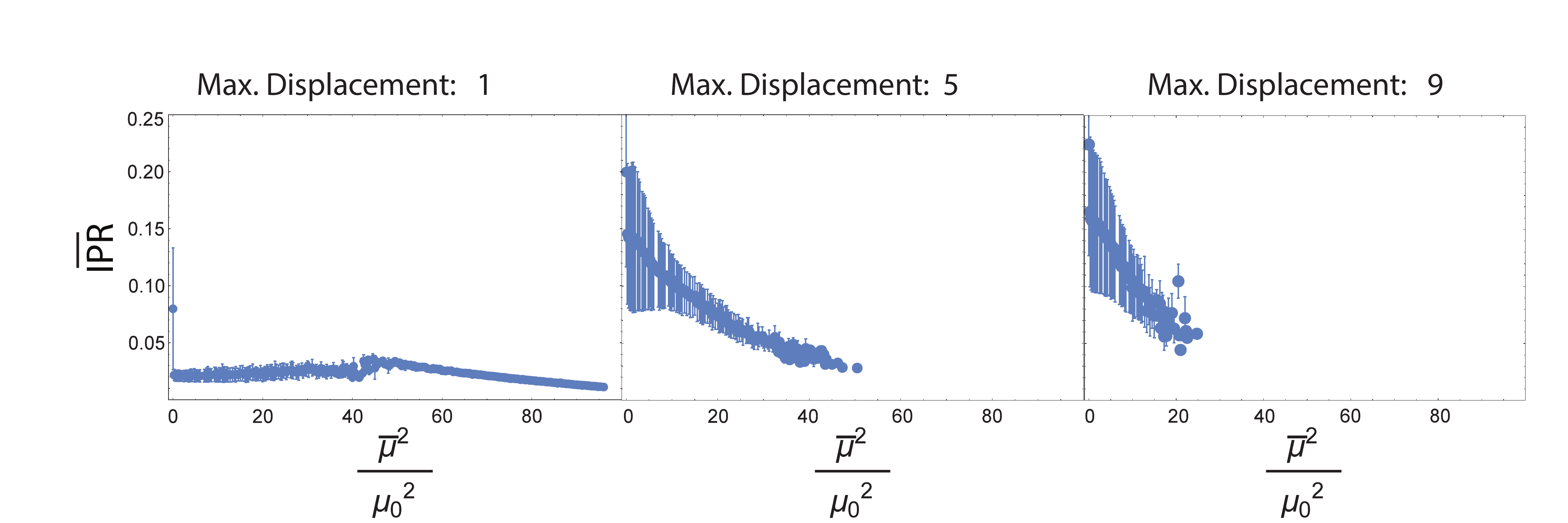}
	\caption{Average IPR as a function of the square dipole strength of 10,000 realizations of  single-chain structures. The error bars indicate the mean absolute deviation from the average number.}
	\label{fig:singlechainipr}
\end{figure}
\begin{figure}
	\centering
		\includegraphics[width=1\textwidth]{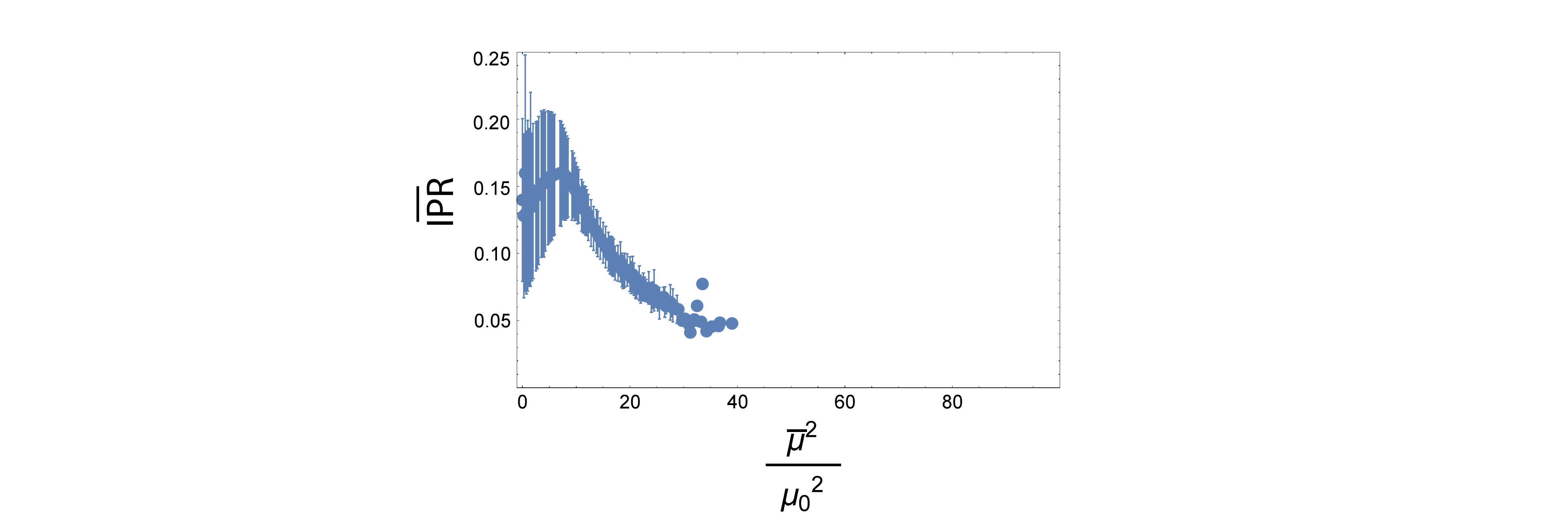}
	\caption{Average IPR as a function of the square dipole strength of 10,000 realizations of  double-chain structures. The error bars indicate the mean absolute deviation from the average number.}
	\label{fig:iprdouble}
\end{figure}
In order to characterize the states delocalization we calculate the inverse participation ratio (IPR), which is a standard measure of delocalization \cite{smyth2012measures}. For an exciton state of the form $|\psi \rangle=\sum_i^N \lambda_i |i\rangle$, the IPR is defined as
\begin{equation}
IPR=\sum_{i=1}^N \lambda_i^4.
\label{eq:ipr}
\end{equation}
The IPR ranges from $1/N$ for a totally delocalized state state to $1$ for a totally localized state.
Below we plot the  IPR as function of the square dipole strength, which determines the oscillator strength for each spectral feature, for different structures  composed of 100 oligomers.  The dipole strength is normalized by the single oligomer dipole strength, therefore it can range from zero for a completely dark state to 100 for a maximally bright state. A dipole strength of one corresponds to a single oligomer. The IPR  ranges from  $1/100$ to 1. Figure~\ref{fig:singlechainipr} shows the average IPR for 10,000 realizations of the single-chain arrangements with a maximum displacements of 1,5 and 9 (from left to right). The error bars indicate the mean absolute deviation from the average.  Figure~\ref{fig:iprdouble} shows the IPR for the double chain structure.  A common feature between all structures is that the brightest states are more delocalized than the dark states. This trend includes the case of the single chain with a maximum monomer displacement of one. Even though its IPR is quite flat compared to the other structures, there is a small slope for the bright states.  As expected, we see also that disorder localizes the states. While the most ordered structure, the single chain with a maximum displacement of 1, has a IPR around 0.02, the others demonstrate IPR values that are much higher.

%

\end{document}